\documentclass[%
 reprint,
superscriptaddress,
showpacs,preprintnumbers,
 amsmath,amssymb,
 aps,
]{revtex4-1}

\usepackage{dcolumn}
\usepackage{bm}
\usepackage[mathlines]{lineno}

\usepackage{amsmath}
\usepackage{units}
\usepackage{graphicx}
\usepackage{dcolumn}
\usepackage{colortbl}
\usepackage{multirow}
\usepackage{graphicx}
\usepackage{booktabs}
\usepackage{array}
\usepackage{ amssymb }
\usepackage{tabularx}
\usepackage[table,xcdraw]{xcolor}
\usepackage{bm}
\usepackage{hyperref}
\usepackage[mathlines]{lineno}

\begin{document}

\title{Overcoming the dephasing limit in multiple-pulse\\ laser wakefield acceleration}

\author{James D. Sadler}
\affiliation{%
Los Alamos National Laboratory, P.O. Box 1663, Los Alamos, New Mexico 87545, USA
}%
\author{Christopher Arran}
\affiliation{%
York Plasma Institute, University of York, York, YO10 5DQ, United Kingdom
}%
\author{Hui Li}
\affiliation{%
Los Alamos National Laboratory, P.O. Box 1663, Los Alamos, New Mexico 87545, USA
}%
\author{Kirk A. Flippo}
\affiliation{%
Los Alamos National Laboratory, P.O. Box 1663, Los Alamos, New Mexico 87545, USA
}%

\begin{abstract}
The electric field in laser-driven plasma wakefield acceleration is orders of magnitude higher than conventional radio-frequency cavities, but the energy gain is limited by dephasing between the ultra-relativistic electron bunch and the wakefield, which travels at the laser group velocity. We present a way to overcome this limit within a single plasma stage. The amplitude of the wakefield behind a train of laser pulses can be controlled in-flight by modulating the density profile. This creates a succession of resonant laser-plasma accelerator sections and non-resonant drift sections, within which the wakefield disappears and the electrons rephase. A two-dimensional particle-in-cell simulation with four $2.5\,$TW laser pulses produces a $50\,$MeV electron energy gain, four times that obtained from a uniform plasma. Although laser red-shift prevents operation in the blowout regime, the technique offers increased energy gain for accelerators limited to the linear regime by the available laser power. This is particularly relevant for laser-plasma x-ray sources capable of operating at high repetition rates, which are highly sought after.
\end{abstract}

\maketitle

The advent of ultra-short high-power lasers has allowed efficient ponderomotive driving of electron plasma waves with phase velocity close to $c$. When the available laser pulses became sufficiently intense, the plasma wave was shown to reach large amplitudes in the non-linear blowout regime \cite{PhysRevLett.43.267, mangles2004monoenergetic, geddes2004high, faure2004laser}. Mono-energetic electron bunches, with charge in the pC range, have been trapped in these waves and accelerated in millimetre size plasmas to energies exceeding $100\,$MeV \cite{mangles2004monoenergetic, leemans2014, gonsalves2019}. These beams are ideal as compact sources for x-ray and gamma photon imaging, demonstrated to have exquisite femtosecond duration and micron source size \cite{fuchs2009laser, powers2014quasi, albert2016applications, PhysRevLett.120.254802, kneip2011x,cole2016, PhysRevX.2.031019}. Compact gamma sources also have potential applications in nuclear material detection. Although the peak x-ray brightness is similar to synchrotron light sources, the average brightness is limited by the $\simeq 1\,$Hz laser firing rate of high-power laser systems, which have low wall-plug efficiencies and struggle with high heat loads. There is therefore considerable interest in using lower peak power lasers with higher repetition rates to accelerate relativistic electrons \cite{goers2015,guenot2017}. This will achieve both high peak brightness and high average brightness in compact future light sources.

However, a severe and inherent limitation of laser plasma acceleration at a given laser power is that the laser pulse travels at a group velocity slightly below $c$, meaning the highly relativistic electrons move forwards relative to the plasma wave and dephase. They eventually enter a part of the wave with a decelerating electric field and lose energy. Dephasing is the primary limitation for accelerators using the linear wakefields produced by lower power lasers; by overcoming the dephasing limit, laser wakefield accelerators could produce electrons with much higher energies without requiring higher laser powers or reducing the repetition rate.

One solution to dephasing is to link several accelerator stages in series, with a new laser pulse driving each of them (e.g. \cite{steinke2016multistage, pousa2019compact, luo2018multistage}). The use of fresh laser pulses avoids complications with laser diffraction and depletion, however it requires coupling every new pulse with femtosecond accuracy. What is more, this approach requires a large increase in the total supplied laser energy, since each new pulse is dumped at the end of the stage. An alternative, more efficient strategy is to manipulate the wakefield using density gradients within a single stage, ensuring that the electron stays in the accelerating region of the wave.

In this work, we propose such a strategy, involving the use of multiple evenly spaced co-linear laser pulses \cite{hooker2014multi}, otherwise known as a resonant laser-plasma accelerator \cite{esarey1996, esarey2009}. The pulse train drives a wakefield only at specific resonant plasma densities, where the pulse spacing is an integer multiple of the plasma wavelength $\lambda_\mathrm{p}$. This density resonance was previously demonstrated experimentally \cite{cowley2017excitation}, using $N=2$ and $N \approx 7$ pulses. By changing the electron density $n_e$ after the dephasing length, the interaction can be moved away from the resonance condition. This allows control over the wakefield amplitude during its propagation. When dephasing occurs, the wakefield can be extinguished, meaning the electrons never encounter a decelerating electric field. The electrons continue to drift and advance relative to the laser pulse, and when they reach the accelerating phase of the wakefield, $n_e$ can be changed back to resonance. The single plasma stage acts in the same way as the multi-stage accelerator technique, with the advantage that no fresh laser pulses are required. 

In practice, the eventual limitation on the scheme is due to the red-shift of the laser pulses from propagating in the wakefield density gradient. This alters their group velocities and affects the pulse spacing. We will show that this limits the parameter window to the linear wakefield regime, reducing the possible output electron energy to far below the multi-GeV level demonstrated in the high-intensity, single-pulse blowout regime \cite{wang2013quasi, leemans2014multi}. However, for accelerators restricted to linear wakefields by the available laser power, the technique offers a significant improvement in energy gain, performing better than the standard scheme with a single laser pulse of equal total energy. This makes the scheme applicable to x-ray photon sources that use linear wakefields excited by high repetition rate terawatt lasers. 

There have been previous proposals for manipulating the wakefield phase using a density ramp \cite{rittershofer2010tapered, PhysRevLett.85.5110, dopp2016energy, malka2016}, shown to increase experimental energy gain by 50\% \cite{PhysRevLett.115.155002}. However, rephasing the electrons in this manner only extends the dephasing limit, rather than overcoming it. Furthermore, use of a density ramp as in Refs. \cite{rittershofer2010tapered, PhysRevLett.85.5110} is limited by group velocity dispersion stretching the laser pulse. It was also shown that density modulations can control the wakefield phase and increase energy gains \cite{yoon2014quasi}. Recently it was shown that, with greater laser energy, advanced focussing techniques can also mitigate dephasing \cite{debus2019circumventing}. These techniques can all produce GeV level energy gains in the high intensity blowout regime. In contrast to these techniques or the multi-stage wakefield accelerators, the scheme we present here is limited to linear wakefields produced by lower power lasers. Although the available energy gains are relatively low, they are several times higher than otherwise possible using the standard single pulse set-up with equal total laser fluence. This allows high-repetition-rate laser systems with low peak powers to produce energetic electrons with greater efficiency.

The novel aspect of our proposal is that use of several laser pulses allows in-flight control over the wakefield amplitude, as well as its phase. For acceleration using linear wakefields, this leads to energy gains several times greater than that from an equivalent uniform plasma. Furthermore, this is the first proposed rephasing scheme that utilises the unique properties of multiple-pulse laser wakefield acceleration. To our knowledge, use of multiple pulses is the only way to gain control over the wakefield amplitude during its propagation. This new aspect of control may lead to innovative electron injection and rephasing schemes, of which this work is a part.

\begin{figure}[t!]
  \centering
  \includegraphics{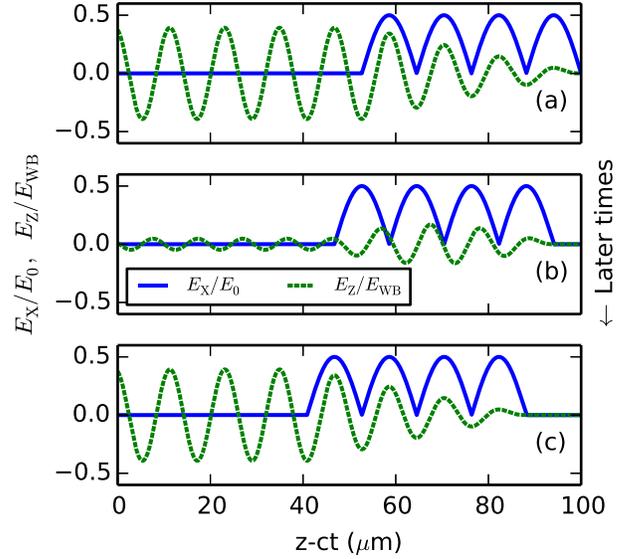}
  \caption{Schematic of the proposed scheme, shown at times (a) $t=0$, (b) $t=L_\mathrm{dp}/c$ and (c) $t=2L_\mathrm{dp}/c$, where $L_\mathrm{dp}$ is the dephasing length. The plots show the laser dimensionless vector potential (solid line), given by the transverse electric field envelope  normalized to $E_0=2\pi m_e c^2/(e\lambda)=12\,$TVm$^{-1}$. Also shown is the longitudinal wakefield electric field (dotted line) normalized to the wave-breaking field $E_\mathrm{WB} = c\sqrt{n_0m_e/\epsilon_0}=270\,$GVm$^{-1}$. There are four evenly spaced laser pulses with peak dimensionless vector potential $a_0=0.5$ and $\lambda = 0.8\,\mu$m, propagating to the right. In panel (b), to prevent deceleration of the electron bunch, the wakefield has been disrupted by raising $n_e$ by 50\%. In panel (c), the bunch is back in phase and so $n_e$ has been returned to $n_0=8\times 10^{18}\,$cm$^{-3}$. }
  \label{fig:schematic}
\end{figure}

Fig. \ref{fig:schematic} illustrates the general scheme. The free electron number density is $n_e\ll n_c$, where the laser critical density is $n_c=\unit[1.1\times10^{21}/\lambda^2]{cm^{-3}}$ and $\lambda$ is the laser wavelength in microns. The wakefield has a wavelength $\lambda_\mathrm{p}=\lambda\sqrt{n_c/n_e}$ and frequency $\omega_\mathrm{p} = 2 \pi c / \lambda_\mathrm{p}$. In each plasma wave period there is a section $\lambda_\mathrm{p}/2$ long where $E_z < 0$ and electrons are accelerated. The wakefield phase velocity equals the laser group velocity $v_g/c=\sqrt{1-n_e/n_c}\simeq 1-n_e/(2n_c)$. The trapped electrons (with velocity $v=\sqrt{1 - \gamma^{-2}}c \simeq c$) dephase when they have travelled a distance $\lambda_\mathrm{p}/2$ relative to the wave. This leads to an expression for the length of acceleration before dephasing occurs, given by $\lambda\left({n_c}/{n_e}\right)^{3/2}$,
which is typically in the millimetre range for $\lambda\simeq1\,\mu$m and $n_e\simeq\unit[10^{18}]{cm^{-3}}$. 

\begin{figure}[t]
  \centering
  \includegraphics{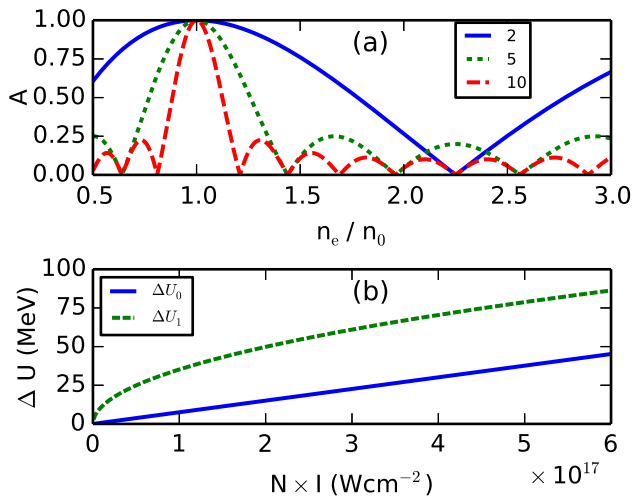}
  \caption{(a) Wakefield amplitude, relative to the value at resonance, for $N=2, 5$ and $10$ laser pulses. This is given by eq. (\ref{eq:amplitude}), as a function of electron density normalised to $n_0=n_c\lambda^2/(c^2\Delta \tau^2)$. (b) Estimates of the maximum achievable electron energy gain for the standard scheme in a uniform plasma ($\Delta U_0$) and for the mitigated dephasing scheme ($\Delta U_1$). This is shown as a function of the laser intensity multiplied by the number of pulses $N$, for laser wavelength $0.8\,\mu$m and $n_0=8\times10^{18}\,$cm$^{-3}$. }
  \label{fig:energy-estimate}
\end{figure}

If the plasma wave is driven by $N$ Gaussian laser pulses, each with temporal intensity profile $I(t) = I_0\exp(-t^2/\sigma^2)$ and spaced by $\Delta\tau$, the wakefield longitudinal electric field amplitude is \cite{cowley2017excitation, gibbon2004short, hooker2014multi}
\begin{align}
E_{z0} = \frac{\sqrt{\pi}m_ecNa_0^2\omega_p^2\sigma}{4e}\exp\left[-\left(\frac{\omega_p\sigma}{2}\right)^2\right]A(n_e, N).\label{EZ}
\end{align}
In this expression $m_e$ is the electron mass, $e$ is the elementary charge and the peak dimensionless vector potential $a_0<1$ is given in terms of laser intensity by $I_0=\unit[1.4\times10^{18}a_0^2/\lambda^2]{Wcm^{-2}}$, for $\lambda$ in units of microns. The resonance function $A$ is given by
\begin{align}
A(n_e, N) = \left|\frac{\sin \left( N \pi \sqrt{{n_e(z)}/{n_0}} \right)  }{ N\sin \left( \pi \sqrt{{n_e(z)}/{n_0}}\right)}\right| ,\label{eq:amplitude}
\end{align}
where $n_0=n_c\lambda^2/(c^2\Delta\tau^2)$ is the resonant plasma density such that $\Delta\tau = \lambda_p/c$. The amplitude $A=1$ when $n_e(z)=n_0$ and $A=0$ when $n_e = n_0 (1 \pm 1/N)^2$, as shown in Fig. \ref{fig:energy-estimate}a.

The phase of the wakefield experienced by an electron bunch travelling at $c$ is described by
\begin{align}
    \phi(z) = \sqrt{\frac{n_e(z)}{n_0}} \left [ \phi_0  + \pi\int_{z_0}^z \frac{n_e(z')}{n_0} ~\frac{\mathrm{d}z'}{L_\mathrm{dp}}  \right ],\label{eq:phase}
\end{align}
where $\phi_0$ is the initial phase of the electron bunch at $z_0$, the dephasing length $L_\mathrm{dp}=\lambda(n_c/n_0)^{3/2}$ and the integral term is due to the varying group velocity of the laser pulse in a density profile. For the case with $n_e=n_0$ and the optimal pulse duration $\sigma = \sqrt{2}/\omega_p$, the maximal electron energy gain over a length $L_\mathrm{dp}$ for $a_0<1$ can be calculated by integrating eq. (\ref{EZ}). The phase in eq. (\ref{eq:phase}) reduces to $\phi = \pi z/L_\mathrm{dp}$, giving the energy gain
\begin{align}
\Delta U_0 &= e\int_0^{L_\mathrm{dp}}E_{z0}\sin(\pi z/L_\mathrm{dp})\,dz \\
&\simeq 1.5m_ec^2Na_0^2n_c/n_0.
\end{align}

As the accelerating field experienced by the electron bunch is $E_z(z) \propto -A(z) \sin [ \phi(z) ]$, the density profile should be designed to maximise $A(z)$ while $0 \leq \phi(z) ~\mathrm{mod}~ 2\pi < \pi$ and minimise $A(z)$ while $\pi \leq \phi(z) ~\mathrm{mod}~ 2\pi < 2\pi$. Note that as the radial electric field is described by $E_r(z,r) \propto - A(z) r \cos [ \phi(z) ]$, this density profile results in the electron bunch experiencing both focussing and defocussing regions of the wakefield. Surprisingly, we found that this reduced the transverse divergence of the electron bunch when compared to a purely focussing scheme, where $A(z)$ was only maximized while $\pi/2 \leq \phi(z) ~\mathrm{mod}~ 2\pi < \pi$. In our scheme the transverse electric field is analogous to the strong focussing regime in conventional particle accelerators, where a periodic arrangement of focussing and defocussing quadrupoles is used to prevent growth in the beam divergence.

Fig. \ref{fig:schematic}b shows the situation after propagation beyond the dephasing length, where $\phi(z) = \phi_0 + \pi$. The density has been increased by $50\%$, so resonance is lost and there is no wakefield during the decelerating section. Due to the large extinction ratio of eq. (\ref{eq:amplitude}) with $N\gg 1$, the exact density profile and its amplitude are not important, only that the increased density $n_1 > n_0(1+\frac{1}{N})^2$. The required length of the acceleration sections is $L_0=L_\mathrm{dp}$ and Eq. (\ref{eq:phase}) shows that the length of the drift sections should be $L_1=L_0n_0/n_1$. In Fig. \ref{fig:schematic}c, $\phi(z)=\phi_0+2\pi$ and the electron bunch is back in phase, albeit in the preceding period. The density has been returned to the resonant value $n_0$.

 \begin{figure}
  \centering
  \includegraphics{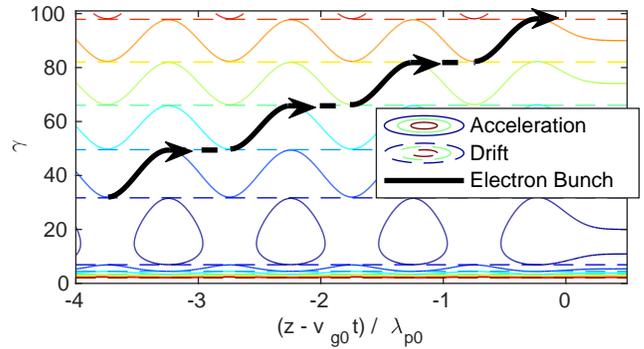}
  \caption{Contours of the Hamiltonian $H(\gamma,z-v_gt)$ for an electron with energy $\gamma m_ec^2$ in the plasma wakefield, shown for the acceleration (solid contours) and drift (dashed) sections of the scheme. The black line shows the trajectory of an example electron bunch. In this example, using four laser pulses each with $a_0 = 0.1$ in plasma at $n_0 = \unit[8 \times 10^{18}]{cm^{-3}}$, the energy gain for each acceleration section is $\Delta U_0 \simeq 16m_ec^2$ and the total gain is $\Delta U_1\simeq 4\Delta U_0$.}
  \label{fig:Hamiltonian}
\end{figure}

Fig. \ref{fig:Hamiltonian} shows the scheme using contours of the relativistic electron Hamiltonian, giving the phase space trajectories. When the wakefield is switched off, the electron maintains constant energy. This allows jumps between separate trajectories, each with energy gain $\simeq \Delta U_0$. The trajectories shown assume optimal acceleration lengths $L_0$, which tend towards $L_\mathrm{dp}$ for large $\gamma$.

The scheme proposed here only requires a spatial modulation of the plasma density with a period of several millimetres. This should be achievable using, for example, multiple gas jets \cite{malka2016} or capillary waveguides \cite{matlis2016b}. Note that laser wakefield acceleration is also limited by diffraction, since the Rayleigh range $z_\mathrm{R} = \pi w_0^2 / \lambda$ will be less than $L_\mathrm{dp}$ if the laser transverse spot size $w_0\simeq \lambda_\mathrm{p}$.  This means acceleration beyond $L_\mathrm{dp}$ requires an overly wide $w_0$ or a laser wave-guide \cite{spence2001, butler2002, lemos2013a, shalloo2018hydrodynamic}. Second order effects such as depletion \cite{shadwick2009nonlinear} and group velocity dispersion will also affect the driver pulses. This paper only describes a way of overcoming dephasing, upon which laser diffraction and depletion become the primary limitations. 

The depletion is associated with laser red-shift, causing an increase in pulse spacing and loss of resonance. The laser wavelength initially evolves \cite{shadwick2009nonlinear} as $\lambda(t)=\lambda_0(1+ct/L_\mathrm{red})$, where $\lambda_0$ is the initial wavelength and $L_\mathrm{red} \simeq 2L_\mathrm{dp}/( Na_0^2)$ is the red-shifting depletion length for the rear-most pulse. By approximating this for short propagation, its group velocity and position is
\begin{align}
     v_g &= c\left(1-\frac{\lambda^2}{2\lambda_\mathrm{p}^2}\right)\simeq\left(1-\frac{\lambda_0^2}{2\lambda_\mathrm{p}^2}\left(1+\frac{ctN a_0^2}{L_\mathrm{dp}}\right)\right),\\
     z &\simeq z_0 + \left(1-\frac{\lambda_0^2}{2\lambda_\mathrm{p}^2}\right)ct  - \frac{N a_0^2\lambda_0^4}{4\lambda_\mathrm{p}^5}(ct)^2.
\end{align}

The plasma wave resonance is lost when the quadratic term reaches a value of $\lambda_\mathrm{p}/2$. This gives the maximum acceleration length as $L=L_\mathrm{dp}\sqrt{{2}/({Na_0^2})}\ll L_\mathrm{red}$ and, accounting for the drift sections, an energy gain of
 \begin{align}
    \Delta U_{1} &\simeq \frac{3n_c}{2n_0}m_ec^2\sqrt{Na_0^2} = \frac{\Delta U_0}{\sqrt{Na_0^2}}.
 \end{align}
 Effective acceleration beyond the dephasing length therefore requires $\sqrt{Na_0^2}\ll 1$, meaning the driver pulses have sub-relativistic intensity and the wakefield is linear. Fig. \ref{fig:energy-estimate}b shows plots of the estimates $\Delta U_0$ and $\Delta U_1$.

Since the scheme is restricted to linear wakefields generated by laser pulses with $a_0\ll 1$, self-injection will not generally occur. Instead, controlled injection is possible through, for instance, a density downramp at the start of the accelerating section \cite{swanson2017}, or an ionization injection scheme using a separate injection pulse \cite{faure2006controlled,bourgeois2013}. A scheme for controlled injection in a resonant multi-pulse accelerator is described in ref. \cite{tomassini2017}. These will lead to lower beam emittance and energy spread than possible with self-injection from high intensity laser pulses, which is a significant advantage of a multi-pulse scheme. It is not however our intent to explore injection in this paper, instead focussing on the accelerating scheme itself.

To further investigate this scheme, and the laser pulse evolution, we conducted a two-dimensional Cartesian particle-in-cell simulation using the code EPOCH \cite{arber2015contemporary}. Due to the linear wakefield and lack of injection mechanisms or self-focussing, a two-dimensional simulation was judged to accurately represent the interaction. The simulation used a moving spatial domain with velocity $c$, length $110\,\mu$m and width $400\,\mu$m. The grid size was 5040 by 576 cells. There were four identical transform limited Gaussian laser pulses, each with wavelength $0.8\,\mu$m, peak intensity $4\times 10^{16}\,$Wcm$^{-2}$, $a_0=0.14$, full width at half maximum duration $15\,$fs and each spaced by $40\,$fs. The transverse waist was $w_0=58\,\mu$m; this made the Rayleigh length similar to the simulated propagation of $13.4\,$mm, with the position of best focus at $z=6.7\,$mm.

  \begin{figure}[t]
  \centering
  \includegraphics{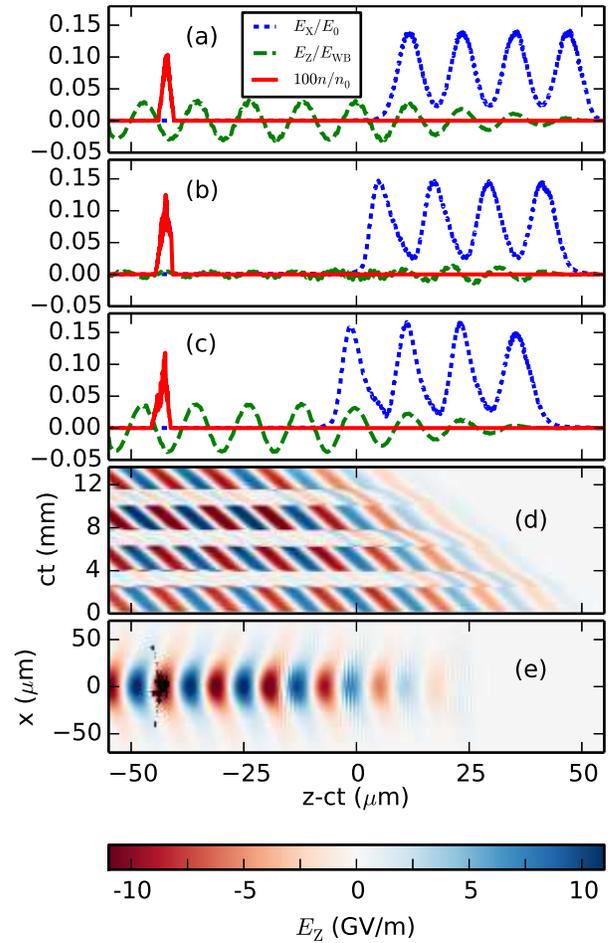}
  \caption{Results of the two-dimensional particle-in-cell simulation, using $n_0=8\times10^{18}\,$cm$^{-3}$, $n_1=1.25\times10^{19}\,$cm$^{-3}$, and four $15\,$fs, $\lambda=0.8\,\mu$m laser pulses with peak intensity $I=4\times10^{16}\,$Wcm$^{-2}$. The panels show axial line-outs at times (a) $ct=1\,$mm, (b) $3\,$mm, and (c) $5\,$mm. The plots show the laser dimensionless vector potential (dotted line), given by the transverse electric field envelope normalized to $E_0=2\pi m_e c^2/(e\lambda)=12\,$TVm$^{-1}$. Also shown is the longitudinal wakefield electric field (dashed line) normalized to the wave-breaking field $E_\mathrm{WB} = c\sqrt{n_0m_e/\epsilon_0}=270\,$GVm$^{-1}$. The relativistic electron density (solid line) is normalized to $n_0$ and multiplied by $100$ for visibility. (d) The on-axis longitudinal electric field across the space-time diagram. (e) The longitudinal electric field at $ct=9\,$mm. Relativistic electrons are shown in black. }
  \label{fig:simulation}
\end{figure}

The resonant plasma density was $n_0=8\times10^{18}\,$cm$^{-3}$, giving $\lambda_\mathrm{p}=11.8\,\mu$m. The density was modulated longitudinally so that the plasma wave lost resonance when the trapped electrons were in a decelerating part of the plasma wave, using $n_1=1.25\times10^{19}\,$cm$^{-3}$. This required repeating a series of acceleration stages of length $L_0=2.3\,$mm and drift stages of length $L_0n_0/n_1=1.5\,$mm. Due to laser pulse evolution, the optimal value of $L_0$ was slightly shorter than $L_\mathrm{dp}=2.6\,$mm. The density transitions had a length-scale of $0.2\,$mm. There were 64 electron macro-particles per cell at temperature $100\,$eV, with static ions. Relativistic electrons were initialised in a $0.2\,$pC bunch in the final plasma wave period in the simulation domain. They had momentum $p_z=15\,$MeV/c, and a Gaussian profile with transverse waist $5\,\mu$m and longitudinal waist $1\,\mu$m. The simulation used the standard Yee field solver with an increased time-step to give accurate numerical dispersion and group velocity. The transverse boundary conditions were periodic for the fields and open for the electron bunch. The standard particle push and current deposition were used \cite{arber2015contemporary}. The simulation set-up is provided as supplementary material.

Fig. \ref{fig:simulation} shows three time-steps from the two-dimensional particle-in-cell simulation, as a function of the co-moving coordinate $z-ct$. In the first time-step, the plasma density is close to the resonant value, giving $\lambda_\mathrm{p}=c\Delta \tau$ and a high amplitude wakefield. Notice how the wakefield amplitude grows linearly with each laser pulse.

\begin{figure}[t]
  \centering
  \includegraphics{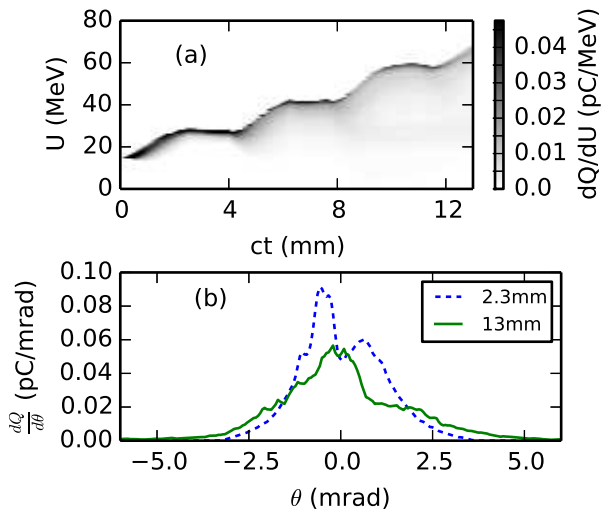}
  \caption{ Results of the two-dimensional particle-in-cell simulation. (a) Spectrum of the electron bunch as a function of time. (b) Angular distribution of the accelerated electrons in the $x$-$z$ plane, shown after the first dephasing length at $ct=2.3\,$mm and at the end of the simulation.}
  \label{fig:simulation-spectrum}
\end{figure}

At the later time-step, shown in Fig. \ref{fig:simulation}b, the relativistic electrons are in the decelerating phase and so the density has been moved away from resonance, causing a loss of wakefield amplitude and minimal deceleration. In Fig. \ref{fig:simulation}c the bunch is again in phase and so the density has been returned to $n_0$ to restore the wakefield, giving another similar burst of acceleration. The laser pulse red-shift and steepening also become apparent at this time. This can also be seen in Fig. \ref{fig:simulation}d, which shows the longitudinal electric field. In the co-moving coordinate, the wakefield steadily shifts backwards due to the laser group velocity being less than $c$. Notice how the wakefield amplitude is effectively extinguished when it would be decelerating the electron bunch. At later times, the extinction ratio in the drift sections becomes weaker. This is because the laser pulse envelopes and spacing have changed. In addition, the longitudinal electric field becomes slightly greater at later times due to the laser pulse steepening, as well as the transverse focussing.

The laser pulse red-shift increases with intensity, meaning the on-axis region of the laser pulse sustains a slower group velocity. This leads to a forwards curvature of the wakefield wave-fronts [Fig. \ref{fig:simulation}e] developing over time.

Some values of $z-ct$ have consistently negative $E_z$ over many dephasing lengths, suggesting that trapped electrons will exceed the dephasing limited energy gain. This is shown by Fig. \ref{fig:simulation-spectrum}a.  The electron bunch retains fairly narrow energy spread, and continues on to receive an energy gain of $\Delta U_1=50\,$MeV, over four times the gain from the first dephasing length $\Delta U_0=12\,$MeV. These values are similar to the estimates in Fig. \ref{fig:energy-estimate}b for the simulated value of $NI=1.6\times 10^{17}\,$Wcm$^{-2}$. 

Emittance growth during the drift stages may also be of concern. However, these sections may be significantly shortened by using a higher density $n_1$. Fig. \ref{fig:simulation-spectrum}b compares the angular beam distribution with respect to the $z$ axis at the end of the simulation and after the first dephasing length. The lengthened propagation does not significantly increase the beam divergence, and only a small fraction of charge escapes the wakefield transversely. 

Motion of the ions presents an additional consideration. However, in reference \cite{hooker2014multi}, it was found that this only becomes significant for $N>50$. Since the plasma density must be maintained within a range of $n_0(1\pm1/N)^2$, it is likely that $N\simeq 5-10$ pulses will be optimal, given experimental engineering constraints on the density uniformity. As the dephasing limit is one of the primary considerations pushing laser wakefield accelerators to lower densities and longer acceleration stages, an additional benefit to eradicating dephasing is that higher plasma densities and shorter propagation stages can be used. This both decreases the laser power requirement for relativistic self-guiding and reduces the need for complex guiding structures. 

Achieving maximal energy transfer efficiency requires a high bunch charge and heavy beam loading, which in this case would cause excessive electron deceleration and spectral dispersion in the drift sections. 

In summary, we have presented a new scheme to overcome the laser-wakefield dephasing limit in the linear regime. With use of several driver pulses, varying the plasma density allows a high degree of control over the wakefield amplitude. With a small change in plasma density, the wakefield can be extinguished, preventing deceleration of the electron bunch. Since the subsequent limitation is depletion and red-shift of the pulses, which does not happen for several dephasing lengths, this allows energy gains many times higher than usual. Furthermore, a two-dimensional simulation demonstrated that the low divergence angle and narrow energy spread is maintained. This work paves the way for use of lower power, kHz repetition rate lasers for generation of relativistic electron beams and laser-plasma photon sources.

Research presented in this article was supported by the Laboratory Directed Research and Development program of Los Alamos National Laboratory under project number 20180040DR. The authors wish to thank the scientific computing staff at Los Alamos National Laboratory for provision of the Grizzly cluster. This work used the EPOCH particle-in-cell code, which was funded in part by the UK EPSRC grants EP/G054950/1, EP/G056803/1, EP/G055165/1 and EP/ M022463/1.





\end{document}